\documentclass[graybox]{svmult}

\usepackage{helvet}         % selects Helvetica as sans-serif font
\usepackage{courier}        % selects Courier as typewriter font
\usepackage{type1cm}        % activate if the above 3 fonts are
\usepackage{makeidx}         % allows index generation
\usepackage{graphicx}        % standard LaTeX graphics tool
\usepackage{multicol}        % used for the two-column index
\usepackage[bottom]{footmisc}% places footnotes at page bottom

\usepackage{amssymb}

\usepackage{url}

\makeindex             % used for the subject index
\newcommand{\be}{\begin{equation}}
\newcommand{\ee}{\end{equation}}
\newcommand{\ud}{{\rm d}}

\begin{document}

\title*{Cosmic-ray propagation in molecular clouds}
\author{Marco Padovani and Daniele Galli}
\authorrunning{Marco Padovani and Daniele Galli}
\institute{M. Padovani \at Laboratoire de Radioastronomie Millim\'etrique, UMR 8112 du CNRS, \'Ecole Normale Sup\'erieure et Observatoire de Paris, 24 rue Lhomond, 75231 Paris cedex 05, France \email{padovani@lra.ens.fr}
\and 
D. Galli \at INAF - Osservatorio Astrofisico di Arcetri, Largo E. Fermi 5, 50125 Firenze, Italy \email{galli@arcetri.astro.it}
%\and
%Alfred~E. Glassgold \at Astronomy Department, University of California, Berkeley, CA 94720-3411, USA\\ \email{aglassgold@berkeley.edu}}
}%
\maketitle

%\abstract*{Each chapter should be preceded by an abstract (10--15 lines long) that summarises the content. The abstract will appear \textit{online} at \url{www.SpringerLink.com} and be available with unrestricted access. This allows unregistered users to read the abstract as a teaser for the complete chapter. As a general rule the abstracts will not appear in the printed version of your book unless it is the style of your particular book or that of the series to which your book belongs.
%Please use the 'starred' version of the new Springer \texttt{abstract} command for typesetting the text of the online abstracts (cf. source file of this chapter template \texttt{abstract}) and include them with the source files of your manuscript. Use the plain \texttt{abstract} command if the abstract is also to appear in the printed version of the book.}

\abstract{Cosmic-rays constitute the main ionising and heating agent in dense,
starless, molecular cloud cores.
We reexamine the physical quantities necessary to determine the cosmic-ray
ionisation rate (especially the cosmic ray spectrum at $E < 1$~GeV and the
ionisation cross sections), and calculate the ionisation rate as a function
of the column density of molecular hydrogen.
Available data support the existence of a low-energy component (below $\sim100$~MeV) of 
cosmic-ray electrons or protons responsible for the ionisation of diffuse and dense clouds.
We also compute the attenuation of the cosmic-ray flux rate in a cloud
core taking into account magnetic focusing and magnetic mirroring, following the propagation
of cosmic rays along flux tubes enclosing different amount of mass and mass-to-flux ratios.
We find that mirroring always dominates over focusing, implying a reduction of the
cosmic-ray ionisation rate by a factor of $3-4$ depending on the position inside the
core and the magnetisation of the core.
}

\section{Introduction}
\label{introduction}
Although the origin of the cosmic radiation and the mechanisms of acceleration are not
well established, this corpuscular radiation has a central role in the high-energy
processes of our galaxy. For instance, cosmic rays (CRs) approximately constitute one third
of the energy density of the interstellar medium and, on a galactic scale, they form
a relativistic gas whose pressure is comparable with that of the galactic
magnetic field.
CR electrons are a source of brem{\ss}trahlung, synchrotron emission and
inverse Compton scattering, revealing the large-scale configuration and the intensity
of the galactic magnetic field. CR nuclei are the unique sample of interstellar
matter which is measurable out of the solar system, 
including all the elements, from hydrogen to actinides.
The abundances of elements and isotopes of CR nuclei bring information not only
with respect to their origin, but, through the radioactive species, show the required
time for their acceleration and the history of their propagation in the interstellar
magnetic field.

Low-energy CRs are the dominant source of ionisation for molecular cloud cores and they 
represent an important source of heating because the energy of primary and secondary 
electrons produced by the ionisation process is in large part converted into heat
by inelastic collisions with ISM atoms and molecules. In addition, the observed diffuse
gamma-ray emission from the galactic plane is believed to be the result of the decay of
neutral pions produced during inelastic collisions of high-energy ($>1$~GeV) CRs with the
ISM~\cite{gab07}.

In general, the CR ionisation rate in the interstellar gas depends on
the relative amount of H, H$_2$, and He~\cite{dyl99}.
The first theoretical determination of the CR ionisation rate was
performed for clouds made only by atomic hydrogen by Hayakawa et al.~\cite{hnt61}.
They assumed
a proton specific
intensity (hereafter, for simplicity, {\em spectrum}) proportional to
the proton energy $E_{\rm p}$ for $0.1~{\rm MeV} < E_{\rm p} < 10~{\rm MeV}$ and
computed $\zeta^{\rm H}\approx 4\times 10^{-16}$~s$^{-1}$.  Spitzer \&
Tomasko~\cite{st68} determined a value (actually a lower limit) of
$\zeta^{\rm H}\gtrsim 6.8\times 10^{-18}$~s$^{-1}$ for \mbox{H\,\textsc{i}} clouds,
assuming a CR proton spectrum declining below $E_{\rm p}\approx 50$~MeV, and
an upper limit of $\zeta^{{\rm H}} \lesssim 1.2\times
10^{-15}$~s$^{-1}$, taking into account an additional flux of $\sim
2$~MeV protons produced by supernova explosions. To obtain the CR
ionisation rate of molecular hydrogen, $\zeta^{{\rm H}_2}$, a useful
approximation is $1.5\zeta^{{\rm H}_2}\approx 2.3\zeta^{{\rm H}}$~\cite{gl74}, 
giving $\zeta^{{\rm H}_2}\approx
10^{-17}$~s$^{-1}$, in agreement with the lower limit on 
$\zeta^{\rm H}$ of Spitzer \& Tomasko~\cite{st68}. This value of 
$\zeta^{{\rm H}_2}$ is often referred to the ``standard'' CR 
ionisation rate in molecular clouds.

A major problem in determining the CR ionisation rate is that low-energy CRs are prevented
from entering the heliosphere by the solar wind and the interplanetary magnetic field.
In fact, the 
outer part of the solar atmosphere, namely the solar corona, is in continuous expansion, 
producing a plasma flux towards the interplanetary medium: the solar wind.
The solar magnetic field is frozen with the plasma and radially dragged 
outwards, and this field 
strongly affects the extrasolar CR particles. In particular, less energetic particles 
are swept away from the solar system and for this reason, the extrasolar CR density 
we observe from the Earth is lower than the density of the local interstellar medium and the 
observed spectra are {\it modulated}, that is altered with respect to their original shapes.
In
practice, this means that Earth-based measurements of CR fluxes give no information on
the interstellar spectrum of protons and heavy nuclei for energies
below $\sim 1$~GeV/nucleon.  Solar modulation also suppresses the flux
of low-energy CR electrons, that already shows considerable fluctuations 
at energies of 10--100~GeV~\cite{cb04}.
Since the cross section for ionisation of molecular hydrogen by
collisions with protons and electrons has a maximum at $\sim 10$~keV
and $\sim 50$~eV, respectively (see Sect.~\ref{reactions_h2}), it is
clear that a knowledge of CR spectrum at low energies is an important
limiting factor for any accurate calculation of the ionisation rate in
the ISM. A direct measurement of the shape of the CR spectrum at these
energies will be possible only when spacecrafts such as {\em
Pioneer} and {\em Voyager} are well beyond the heliopause, the
outermost boundary for solar modulation effects, believed to lie at
100--150~AU from the Sun.

Over the past three decades, several values of $\zeta^{\rm H}$ ranging
from a few $10^{-17}$~s$^{-1}$ to a few $10^{-16}$~s$^{-1}$ have been
obtained in diffuse interstellar clouds from measurements of the
abundances of various chemical species, in particular OH~\cite{bd77,hbd78,bhd78}
and HD~\cite{vdb86,fwl96}. However, the derived rates are sensitive to
several model assumptions, e.g. the value of specific chemical reaction
rates and the intensity of the UV background. In dense molecular
clouds, determining the CR ionisation rate is made even more
uncertain by the sensitivity of molecular abundances to the level of
depletion of the various species and the role of small and large grains
in the chemical network. The values of $\zeta^{{\rm H}_2}$ derived by
Caselli et al.~\cite{c98} through DCO$^+$ and
HCO$^+$ abundance ratios span a range of about two orders of magnitudes
from $\sim 10^{-17}$~s$^{-1}$ to $\sim 10^{-15}$~s$^{-1}$, with a
scatter that may in part reflect intrinsic variations of the CR flux
from core to core. Finally, values of $\zeta^{{\rm H}_2}$ of a few
times $10^{-17}$~s$^{-1}$ have been obtained in clouds of higher column
density ($N({\rm H}_2)\sim 10^{23}$--$10^{24}$~cm$^{-2}$) like the
envelopes surrounding massive protostellar sources~\cite{vdtvd00,d02}.

The discovery of significant abundances of H$_3^+$ in diffuse clouds~\cite{mcc98}, 
confirmed by follow-up detections~\cite{g99,mcc03,i12}, has led to values
of $\zeta^{{\rm H}_2}$ larger by about one order of magnitude than both
the ``standard'' rate and previous estimates based on the abundance of
OH and HD in dense clouds. Given the relative simplicity of the
chemistry of H$_3^+$, it is now believed that diffuse clouds are
characterised by CR ionisation rates $\zeta^{{\rm H}_2}\approx 2\times
10^{-16}$~s$^{-1}$ or larger. This value has been confirmed by
Neufeld et al.~\cite{neu10} who found $\zeta^{{\rm H}_2}=0.6-2.4\times10^{-16}$ s$^{-1}$
probing
the CR ionisation rate in clouds of low H$_{2}$ fraction from observations of
OH$^{+}$ and H$_{2}$O$^{+}$.

\section{CR reactions with H$_2$ and He}
\label{reactions_h2}

CR particles (electrons, protons, and heavy nuclei) impact with atoms and
molecules of the ISM producing ions and electrons. Table~\ref{react} lists
the main CR ionisation reactions involving H$_2$ and He.  

\begin{table}[!h]
\begin{center}
\caption{CR reactions in molecular clouds.}
\begin{tabular}{lll}
\hline\hline
$p_{\rm CR} + {\rm H}_2 \rightarrow p_{\rm CR} + {\rm H}_2^+ + e$          &\hspace{1cm} &  $\sigma_p^{\rm ion.}$\\
$p_{\rm CR}+{\rm H}_2 \rightarrow {\rm H} + {\rm H}_2^+$                   &&  $\sigma_p^{\rm e.~c.}$\\
$p_{\rm CR} + {\rm H}_2 \rightarrow p_{\rm CR} + {\rm H} + {\rm H}^+ + e$  &&  $\sigma_p^{\rm diss.~ion.}$\\
$p_{\rm CR} + {\rm H}_2 \rightarrow p_{\rm CR} + 2{\rm H}^+ + 2e$          &&  $\sigma_p^{\rm doub.~ion.}$\\
\hline
$e_{\rm CR} + {\rm H}_2 \rightarrow e_{\rm CR} + {\rm H}_2^+ + e$          &&  $\sigma_e^{\rm ion.}$\\
$e_{\rm CR} + {\rm H}_2 \rightarrow e_{\rm CR} + {\rm H} + {\rm H}^+ + e$  &&  $\sigma_e^{\rm diss.~ion.}$\\
$e_{\rm CR} + {\rm H}_2 \rightarrow e_{\rm CR} + 2{\rm H}^+ + 2e$          &&  $\sigma_e^{\rm doub.~ion.}$\\
\hline
$p_{\rm CR} + {\rm He} \rightarrow p_{\rm CR} + {\rm He}^+ + e$            &&  $\sigma_p^{\rm ion.}$\\
$p_{\rm CR}+{\rm He} \rightarrow {\rm H} + {\rm He}^{+}$                      &&  $\sigma_p^{\rm e.~c.}$\\
\hline
$e_{\rm CR} + {\rm He} \rightarrow e_{\rm CR} + {\rm He}^+ + e$            &&  $\sigma_e^{\rm ion.}$\\
\hline
\end{tabular}
\label{react}
\end{center}
\end{table}

In molecular
clouds, a large majority of CR--H$_2$ impacts leads to the formation
of H$_2^+$ via the {\em ionisation} reaction
\be
k_{\rm CR} + {\rm H}_2 \rightarrow k_{\rm CR} + {\rm H}_2^+ + e,
\label{ion}
\ee
where $k_{\rm CR}$ is a cosmic-ray particle of species $k$ and energy
$E_k$, with cross section $\sigma_k^{\rm ion.}$.  Here we consider CR
electrons ($k=e$), protons ($k=p$), and heavy nuclei of charge $Ze$
($k=Z$, with $Z \ge 2$).  Low-energy CR protons, in addition, may react
with ambient H$_2$ by {\em electron capture} reactions,
\be
p_{\rm CR}+{\rm H}_2 \rightarrow {\rm H} + {\rm H}_2^+,
\label{p_ec}
\ee
with cross section $\sigma_p^{\rm e.~c.}$.
For an isotropic distribution of CR particles, the production rate of
H$_2^+$ (per H$_2$ molecule) is then
\begin{eqnarray}
\zeta^{{\rm H}_2} &=& 4\pi \sum_k
\int_{I({\rm H}_2)}^{E_{\rm max}} j_k(E_k)[1+\phi_k(E_k)]
\sigma_k^{\rm ion.}(E_k)\,\ud E_k \nonumber \\
&&+ 4\pi\int_{0}^{E_{\rm max}} j_p(E)
\sigma_p^{\rm e.~c.}(E_p)\,\ud E_p, 
\label{zetatot_h2}
\end{eqnarray}
where $j_k(E_k)$ is the number of CR particles of species $k$ per unit
area, time, solid angle and per energy interval (hereafter, we will
refer to $j_k(E_k)$ simply as the spectrum of particle $k$), $I({\rm
H}_2)=15.603$~eV is the ionisation potential of H$_2$, and $E_{\rm
max}=10$~GeV is the maximum energy considered.

The quantity
$\phi_k(E_k)$ is a correction factor accounting for the ionisation of
H$_2$ by secondary electrons. In fact, secondary electrons are
sufficiently energetic to induce further ionisations of H$_2$
molecules, and their relatively short range justifies a local treatment
of their ionising effects.  The number of secondary ionisation produced
per primary ionisation of H$_2$ by a particle $k$ is determined by
\be\label{secondel}
\phi_k(E_k)\equiv \frac{1}{\sigma_k^{\rm ion.}(E_k)}
\int_{I({\rm H}_2)}^{E^\prime_{\rm max}}P(E_k,E_e^\prime)
\sigma^{\rm ion.}_e(E_e^\prime)\,\ud E_e^\prime,
\ee
where $P(E_k,E_e^\prime)$ is the probability that a secondary electron
of energy $E_e^\prime$ is ejected in a primary ionisation by a particle
of energy $E_k$. The spectrum of secondary electrons declines rapidly
with $E_e^\prime$ from the maximum at $E_e^\prime=0$ (Glassgold \&
Langer~\cite{gl73b}; Cecchi-Pestellini \& Aiello~\cite{cpa92}).  The function
$\phi_e(E_e)$ giving the number of secondary ionisations after a single
ionisation by an electron of energy $E_e$ has been computed by
Glassgold \& Langer~\cite{gl73b} for energies of the incident electron up to
10~keV. Above a few 100~eV, $\phi_e$ increases logarithmically with
$E_e$. For secondary electrons produced by impact of particles $k$, we
adopt the scaling $\phi_k(E_k)\approx \phi_e(E_e=m_e E_k/m_k)$ valid in
the Bethe-Born approximation. Calculations by Cravens \&
Dalgarno~\cite{cd78} confirm this scaling for protons in the range
1--100~MeV.

Additional ionisation reactions that produce electrons are
the {\em dissociative ionisation} of H$_2$ and the {\em double ionisation} of H$_2$.
These two processes contribute 
to the total CR production rate of electrons per H$_2$ molecule,
\begin{eqnarray}
\zeta^e&=&4\pi \sum_k
\int_{I({\rm H}_2)}^{E_{\rm max}} j_k(E_k)[1+\phi_k(E_k)]
\sigma_k^{\rm ion.}(E_k)\,\ud E_k \nonumber \\
&&+ 
4\pi \sum_k\int_{E^{\rm diss.~ion.}}^{E_{\rm max}} j_k(E)[1+\phi_k(E_k)]
\sigma_k^{\rm diss.~ion.}(E_k)\,\ud E_k \nonumber \\
& &+ 
8\pi \sum_k\int_{E^{\rm doub.~ion.}}^{E_{\rm max}} j_k(E_k)[1+\phi_k(E_k)]
\sigma_k^{\rm doub.~ion.}(E_k)\,\ud E_k\ ,
\label{zetatot_e}
\end{eqnarray}
where $E^{\rm diss.~ion.}=18.1$~eV and $E^{\rm doub.~ion.}=51$~eV.
The cross sections of these processes are smaller by at
least one order of magnitude than the corresponding ionisation cross
section, and the relative contribution of dissociative ionisation and
double ionisation to the total electron production rate is expected to
be small~\cite{pgg09}.

Similarly, the CR production rate of He$^+$ (per He atom) is 
\begin{eqnarray}
\zeta^{{\rm He}}&=&4\pi \sum_k
\int_{I({\rm He})}^{E_{\rm max}} j_k(E_k)[1+\phi_k(E_k)]
\sigma_k^{\rm ion.}(E_k)\,\ud E_k \nonumber \\
&& +4\pi\int_{0}^{E_{\rm max}} j_p(E)\sigma_p^{\rm e.~c.}(E_p)\,\ud E_p\ , 
\label{zetatot_he}
\end{eqnarray}
where $I({\rm He})=24.587$~eV is the ionisation potential of He,
$\sigma_k^{\rm ion.}$ is the ionisation cross sections of He for impact
by particles $k$, and $\sigma_p^{\rm e.~c.}$ is the electron capture
cross section.

\begin{figure}[!h]
\begin{center}
\resizebox{.9\hsize}{!}{\includegraphics{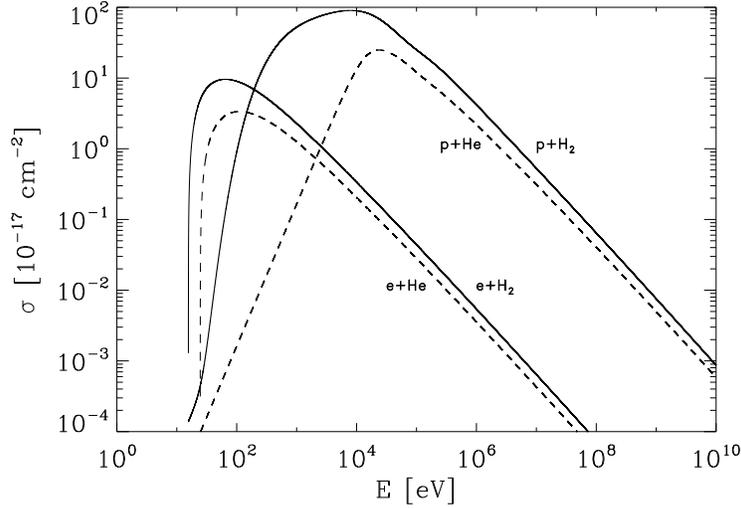}}
\caption[]{Ionisation cross sections for proton and electron impact on H$_{2}$ ({\em solid lines})
and on He ({\em dashed lines}).}
\label{sigma_santcugat}
\end{center}
\end{figure}

Figure~\ref{sigma_santcugat} shows that ionisation cross sections peak at about
10~keV and 0.1~keV for protons and electrons colliding with H$_{2}$, respectively,
and at about 30~keV and 0.1~keV for protons and electrons colliding with He, respectively. 
This means that, to compute reliable CR ionisation rates, the CR spectrum needs to be extrapolated down to $\sim$~keV energies where the ionisation cross sections have their maximum.

\section{Local interstellar spectra}
\label{LIS}
It is generally assumed that the local interstellar (LIS) spectrum, namely the
CR spectrum in the solar neighbourhood, 
characterises the energy
distribution of CR everywhere in the galactic disk, as long as the ISM
properties do not depart from the uniform conditions assumed in the
propagation model. %With this assumption, Webber~(1998) adopted LIS
%spectra for protons and heavy nuclei of energy greater than $10$~MeV
%and electrons of energy greater than 2~MeV and combined them with data
%from {\em Voyager} and {\em Pioneer} spacecraft measurements out to
%60~AU from the Sun to obtain a CR ionisation rate $\zeta^{{\rm
%H}}\approx 3$--$4\times 10^{-17}$~s$^{-1}$.  This is 5--6 times the
%``standard'' rate of Spitzer \& Tomasko~(1968) for atomic hydrogen.

It is very uncertain, however, whether the LIS spectrum is really
representative of the whole galactic disk, especially because the Solar
System resides in a low-density ($n\approx 10^{-3}$~cm$^{-3}$) region.
%produced by $\sim 10$ supernovae exploded over the past $10$~Myr (the
%``Local Bubble''). 
%In addition, to compute reliable CR ionisation
%rates, the demodulated spectra need to be extrapolated down to $\sim
%\mbox{keV}$ energies where the ionisation cross sections have a maximum
%(see Sect.~\ref{reactions_h2}, \ref{reactions_e} and
%\ref{reactions_he}).  Given these uncertainties, we discuss in the
%remainder of the paper the consequences for the CR ionisation rate of
%making different assumptions about the low-energy behaviour of CR
%spectra.  In particular, we consider for both protons and electrons a
%``minimum'' and ``maximum'' LIS spectrum compatible with the available
%observational constraints, and we compute the resulting ionisation
%rates with the objective of comparing them with existing data for
%diffuse and dense clouds.
We assume a uniform distribution
(in space and time) of CR sources characterised by a given ``source
spectrum''. CR propagation models can
generate steady-state LIS spectra resulting from a
number of processes affecting the CR transport in the galactic disk (see Sect.~\ref{Elosses}),
which, in turn, can be used as input for solar modulation
calculations to reproduce the CR spectrum and the relative abundances
of CR particles measured at the Earth.  
The LIS spectra obtained in
this way are clearly not uniquely defined, and a considerable range of
LIS spectral shapes can be shown to be consistent with the measured CR
flux with appropriate choices of parameters of the transport model (see
e.g. Mewaldt et al.~\cite{m04}, especially their Fig.~1).

\begin{figure}[!h]
\begin{center}
\resizebox{\hsize}{!}{\includegraphics{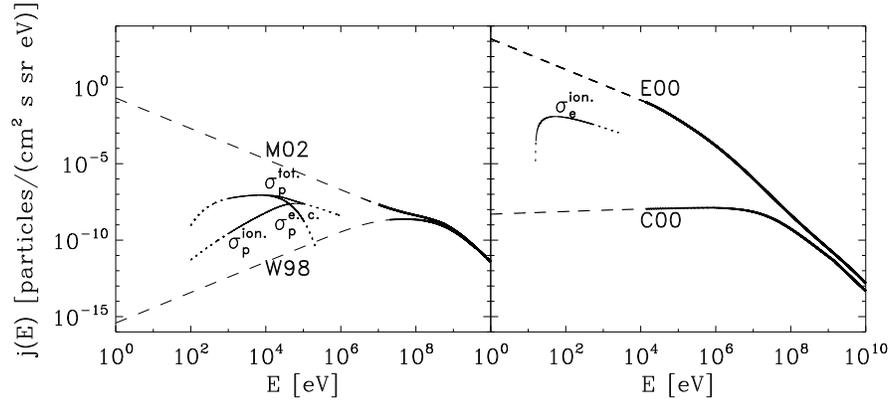}}
\caption[]{{\em Left panel: }proton LIS spectra of M02 and W98 ({\em
solid} curves). The {\em dashed} curves represent our
power-law extrapolations of the spectra. For comparison, the
cross sections for ionisation of H$_2$ by proton impact, electron
capture, and total ionisation are also shown.
{\em Right panel: }electron LIS spectra of E00 and C00 ({\em
solid} curves). The {\em dashed} curves represent our
extrapolations of the spectra. For comparison, the cross section
for ionisation of H$_2$ by electron impact is also shown. All the cross sections are
in
arbitrary units.}
\label{LISspectra}
\end{center}
\end{figure}

We consider for both protons and electrons a
``minimum'' and ``maximum'' LIS spectrum compatible with the available
observational constraints, and we compute the resulting ionisation
rates with the objective of comparing them with existing data for
diffuse and dense clouds.
We extrapolate the LIS spectra for the CR proton and electron components to lower
energies with power-laws to reach the peak of the ionisation cross section.
The two determinations of the proton LIS spectrum (left panel of Fig.~\ref{LISspectra})
that we considered are given by
Webber~\cite{w98} (``minimum'', hereafter W98) 
and Moskalenko et al.~\cite{m02} (``maximum'', hereafter M02).
W98 estimated the LIS proton spectrum down to $\sim$~10~MeV, starting from an injection
spectrum parameterised as a power-law in rigidity, propagated according to the model of
Webber~\cite{w87} and accounting for solar modulation following Potgieter~\cite{p95}.
M02 reproduces the observed spectrum of protons, antiprotons, alpha nuclei, the B/C ratio,
and the diffuse $\gamma$-ray background. 
The two electron LIS spectra adopted (right panel of Fig.~\ref{LISspectra}) both derive by Strong et al.~\cite{smr00}: the former
(``minimum'', hereafter C00) is mostly derived from radio observations. It reproduces the
spectrum of electrons, protons, and alpha nuclei above $\sim$~10~GeV, but fails to account
for the $\gamma$-ray background for photons with energies below $\sim$~30~GeV and above
$\sim$~1~GeV. The latter (``maximum'', hereafter E00) reproduces the $\gamma$ observations
at photon energies below $\sim$~30~GeV by a combination of brem{\ss}trahlung and inverse
Compton emission, assuming a steepening of the electron spectrum below $\sim$~200~MeV to
compensate for the growth of ionisation losses. Our extrapolations at low energies are
power-law in energy, $j(E)\propto E^{\beta}$. The values of $\beta$ are shown in 
Table~\ref{ion_LIS}.

The values of $\zeta_k^{{\rm H}_2}$, $\zeta_k^e$ and $\zeta_k^{{\rm
He}}$ per H$_2$ molecule and He atom, respectively, obtained from
numerical integration of Eq.~(\ref{zetatot_h2}), (\ref{zetatot_e}) and (\ref{zetatot_he}), with the $j_k(E_k)$ taken to be the adopted LIS
spectra, are listed in Table~\ref{ion_LIS}. We have assumed a mixture of
H$_2$ and He with $n(\mathrm{H_{2}})=0.5n(\mathrm{H})$ and
$n(\mathrm{He})=0.1n(\mathrm{H})$, that is $f_{{\rm H}_2}\sim0.83$ and $f_{\rm He}\sim0.17$,
where $f_{k}=n(k)/[n(\mathrm{H_{2}})+n(\mathrm{He})]$, with $k=\mathrm{H_{2}},\mathrm{He}$.  
We checked that our extrapolations at low energies were compatible with energy constraints.
To this purpose, we computed the energy density of each CR component, defined as
\be 
{\cal E}_k=4\pi\int_0^\infty \frac{j_k(E_k)E_k}{v_k(E_k)}\,\ud E_k
\label{en_den}
\ee
where $j_k(E_k)$ is the particle's LIS spectrum and
$v_k(E_k)=c(E_k^2/m_k^2c^4+2E_k/m_kc^2)^{1/2}/(1+E_k/m_kc^2)$ is the
velocity of particle $k$ with kinetic energy $E_k$. We compute the
total energy density of CR as $\sum_k{\cal E}_k\approx (1+\xi) {\cal
E}_p$, where $\xi=0.41$ is the correction factor for the abundance of
He and heavy nuclei. We found that the total CR energy density
varies from a minimum of $0.97$~eV~cm$^{-3}$ (W98 plus C00) and a
maximum of $1.80$~eV~cm$^{-3}$ (M02 plus E00), corresponding to an
equipartition magnetic field of $6.2$~$\mu$G and $8.5$~$\mu$G,
respectively.  These equipartition values are compatible with the
``standard'' value of the magnetic field of $6.0\pm 1.8$~$\mu$G in the
cold neutral medium of the Galaxy (Heiles \& Troland~\cite{ht05}).

\begin{table}[!h]
\begin{center}
\caption{CR ionisation rates $\zeta^{{\rm H}_2}_k$ 
and $\zeta^{\rm He}_k$, %(per H$_2$ and per He, respectively),
electron production rate $\zeta^e_k$, energy densities ${\cal E}_k$, and exponent of the
power-law extrapolation of the spectrum at low energies for CR protons$^{a}$ ($p$) and  electrons ($e$). 
%for the LIS spectra assumed in this work.
}
\begin{tabular}{llllllll}
\hline
$k$\hspace{.2cm}&  ref.\hspace{.2cm}      & $\zeta_k^{{\rm H}_2}$\hspace{.2cm}   & $\zeta_k^{\rm He}$\hspace{.2cm}     & $\zeta_k^e$  & ${\cal E}_k$   & \hspace{.5cm}$\beta$\\
\hspace{.2cm}&\hspace{.2cm}& [s$^{-1}$]              & [s$^{-1}$]             & [s$^{-1}$]   & [eV~cm$^{-3}$] & \hspace{.5cm}\\
\hline\hline
$p$\hspace{.2cm} &      W98 \hspace{.2cm}  & $2.08\times 10^{-17}$\hspace{.2cm}   & $1.33\times 10^{-17}$\hspace{.2cm}  & $2.50\times 10^{-17}\hspace{.2cm}$ & 0.953 & \hspace{.5cm}0.95\\
$p$ &      M02   & $1.48\times 10^{-14}$   & $2.16\times 10^{-15}$  & $3.49\times 10^{-15}$ & 1.23  & \hspace{.5cm}$-1$\\
\hline
$e$   &    C00   & $1.62\times 10^{-19}$   & $1.05\times 10^{-19}$  & $1.94\times 10^{-19}$ & 0.0167 & \hspace{.5cm}0.08\\
$e$   &    E00   & $6.53\times 10^{-12}$   & $2.46\times 10^{-12}$  & $7.45\times 10^{-12}$ & 0.571 & \hspace{.5cm}$-1$\\
\hline
\end{tabular}\\[2pt]
$^a$ The proton ionisation rates include
the contribution of heavy nuclei.
\label{ion_LIS}
\end{center}
\end{table}

\section{Energy losses of CRs in the ISM}
\label{Elosses}
The penetration of primary CR and secondary particles in interstellar clouds was studied by
Takayanagi~\cite{t73} and in more detail by Umebayashi \& Nakano~\cite{un81}. The quantity
which describes the ``degradation spectrum'' of the CR component $k$ 
resulting from the energy loss of the
incident particles is called {\it energy loss function}, defined by
\be
L_k(E_k)=-\frac{1}{n({\rm H}_2)}\left(\frac{\ud E_k}{\ud \ell}\right),
\label{eloss}
\ee
where $n({\rm H}_2)$ is the density of the medium in which the
particles propagate and $\ell$ is the path length. Since we consider
only energy losses in collisions with H$_2$, our results are applicable
only to clouds in which hydrogen is mostly in molecular form.
Some energy loss processes are common to CR protons and electrons, like Coulomb
interactions, inelastic collisions and ionisation; others are peculiar to protons
(elastic collisions, pion production and spallation), others to electrons (brem\ss trahlung,
synchrotron emission and inverse Compton scattering). These processes are briefly reviewed in
the following subsections.

\subsection{Energy loss of protons colliding with H$_2$}
\label{energy_loss_p}

To determine the energy loss function of protons we have used the
results collected by Phelps~\cite{pav90} for energies in the range from
10$^{-1}$~eV to 10$^4$~eV. For higher energies, between 1~keV and
10~GeV, we have used data from the NIST Database\footnote{\tt
http://physics.nist.gov/PhysRefData/Star/Text} for atomic hydrogen
multiplied by a factor of 2 to obtain the corresponding values for
collisions with molecular hydrogen. NIST data do not include pion
production at energies higher than about $0.5$~GeV, that we computed
following Schlickeiser~\cite{s02}. The resulting energy loss function is
shown in the left panel of
Fig.~\ref{Eloss_Range_SantCugat}. The broad peak in $L_p(E_p)$ at
$E_p\approx 10$~eV is due to elastic collisions and to the
excitation of rotational and vibrational levels, the peak at
$E_p\approx 100$~keV to ionisation, and the rapid increase at energies
above $\sim 1$~GeV is due to pion production.  For the low ionisation
levels characteristic of molecular clouds, the energy loss for Coulomb
interactions of CRs with ambient electrons can be neglected at energies
above $\sim 1$~eV (dashed line in the left panel of Fig.~\ref{Eloss_Range_SantCugat}), since 
$L_p(E_p)\propto n_e E_p^{-0.5}$, where $n_e$ is
the electron density of the traversed matter~\cite{s02}.

\subsection{Energy loss of electrons colliding with H$_2$}
\label{energy_loss_e}

\begin{figure}[!h]
\begin{center}
\resizebox{\hsize}{!}{\includegraphics{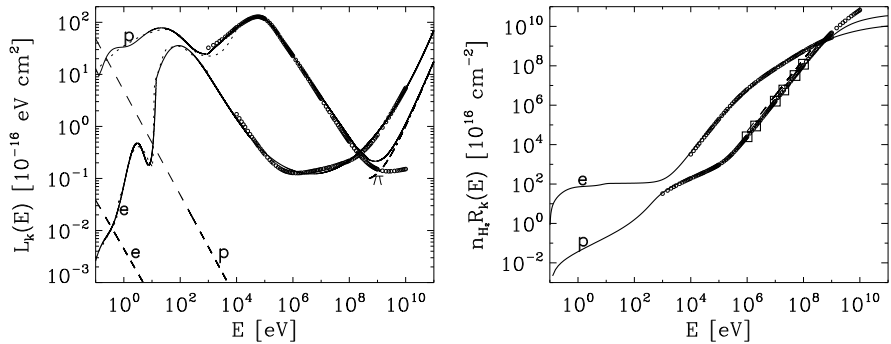}}
\caption[]{{\em Left panel:} 
energy loss functions $L_e(E_e)$ and $L_p(E_p)$ for electrons
and protons, respectively, colliding with H$_2$ ({\em solid} curves), compared with
NIST data ({\em circles}); {\em dashed} curves show Coulomb losses for 
a fractional electron abundance $n_e/n({\rm H}_2)=10^{-7}$; {\em
dash-dotted} curve labeled with $\pi$ represents the energy loss by pion
production computed following Schlickeiser~\cite{s02}; {\em dotted} curves show
the results by Phelps~\cite{pav90} and Dalgarno et al.~\cite{dyl99} for $p$--H$_2$
and $e$--H$_2$, respectively. {\em Right panel:} 
range $R_e(E_e)$ and $R_p(E_p)$ for electrons and protons
colliding with H$_2$ ({\em solid} curves), respectively, compared with NIST data ({\em
circles}) and the results of Cravens \& Dalgarno~\cite{cd78}, {\em squares};
the {\em dashed} curve shows the fit by Takayanagi~\cite{t73}.}
\label{Eloss_Range_SantCugat}
\end{center}
\end{figure}

To determine the electron energy loss function we have adopted the
results of Dalgarno et al.~\cite{dyl99} for $10^{-2}~{\rm eV} \le E_e \le
1~{\rm keV}$ and those of Cravens, Victor \& Dalgarno~\cite{cvd75} for
$1~{\rm eV} \le E_e \le 10$~keV. For higher energies, $10~{\rm keV} \le
E_e \le 10~{\rm GeV}$, we have adopted the loss function for electron-H
collisions from the NIST Database multiplied by a factor of 2. The
resulting energy loss function is also shown in
the left panel of Fig.~\ref{Eloss_Range_SantCugat}. The first peak in $L_e(E_e)$
is due to the
excitation of vibrational levels, the second to the excitation of the
electronic levels and ionisation, while at higher energies the energy
loss function is dominated by brem\ss trahlung.  As in the case of CR
protons, we can neglect the contribution of Coulomb interactions for
electrons at energies above $\sim 1$~eV, since 
$L_e(E_e)\propto n_e^{0.97}E_p^{-0.44}$~\cite{sng71}.  

%In Fig.~\ref{Figure8}, we show
%the range for electrons in H$_2$, obtained as in the case of CR
%protons, compared with data from the NIST Database for $10~{\rm keV}
%\le E_{\rm e} \le 1~{\rm GeV}$.
\section{Propagation of a cosmic ray in a molecular cloud}
\label{propagation}

We assume a plane-parallel geometry and we follow the propagation of CR particles inside
a molecular cloud with the so-called {\it continuous-slowing-down approximation} (hereafter
CSDA) which is
also referred as the {\it continuous energy loss regime} or {\it thick target
approximation} when the propagation is dominated by these losses~\cite{rl75,rl96}.
It is useful to introduce the column density of molecular hydrogen $N$(H$_{2}$),
\be
N({\rm H}_2)=\int n({\rm H}_2)\,\ud \ell\ ,
\ee
and to rewrite the energy loss function (Eq.~\ref{eloss}) as
\be
L_k(E_k)=-\frac{\ud E_k}{\ud N({\rm H}_2)}\ .
\label{L(E)N}
\ee
Let us then define $j_k(E_k,N)$ as the spectrum of CR particles of
species $k$ at depth $N({\rm H}_2)$, with $j_k(E_k,0)$ representing
the LIS spectrum incident on the cloud's surface, defined by a column
density $N({\rm H}_2)=0$. To compute $j_k(E_k,N)$ we must consider all
the processes that degrade the energy of the incident CR particles.
Assuming that the direction of propagation does not change significantly
inside the cloud, it follows from Eq.~(\ref{L(E)N}) that particles of
initial energy $E_{k,0}$ reach energy $E_k<E_{k,0}$ as a consequence of
energy losses after propagating across a column density $N({\rm H}_2)$
given by
\be
N({\rm H}_2)=-\int_{E_{k,0}}^{E_k}\frac{\ud E_k}{L_k(E_k)} 
=n({\rm H}_2)[R_k(E_{k,0})-R_k(E_k)]\ ,
\label{defN}
\ee
where $R_k(E_k)$ is the {\em range}, defined as
\be
R_k(E_k)=\int_{E_k}^0\ud \ell=
\int_0^{E_k}\frac{\ud E_k}{-(\ud E_k/\ud \ell)}=
\frac{1}{n({\rm H}_2)}\int_0^{E_k}\frac{\ud E_k}{L_k(E_k)}\ .
\label{defrange}
\ee
In the right panel of 
Fig.~\ref{Eloss_Range_SantCugat} 
we show the quantity $n({\rm H}_2)R_k(E_k)$ for $k=p,e$, 
obtained with a numerical integration of Eq.~(\ref{defrange}), compared with data
from the NIST Database at energies from 1~keV to 10~GeV for protons and from
10~keV to 1~GeV for electrons. 
For protons we also show the
fit adopted by Takayanagi~\cite{t73} in a limited range of energies and the
results of Cravens \& Dalgarno~\cite{cd78}. As one can see, except for energies
higher than $\sim 100$~MeV, where the NIST data do not include energy
losses by pion production, there is a complete agreement between our results and the
NIST data.

Conservation of the number of CR particles of each species implies 
\be
j_k(E_k,N)\,\ud E_k=j(E_{k,0},0)\,\ud E_{k,0}\ ,
\ee
where, for a given value of $N({\rm H}_2)$, the infinitesimal variation $\ud
E_{k,0}$ of the particle's initial energy corresponds to an infinitesimal
variation $\ud E_k$ of its energy at a depth $N({\rm H}_2)$ given by
\be
\frac{\ud E_k}{L_k(E_k)}=\frac{\ud E_{k,0}}{L_k(E_{k,0})}
\label{jacob}
\ee
We ignore here that electron capture reactions of CR protons with H$_2$ and He
do not conserve the number of CR protons and also the 
$\alpha + \alpha$ fusion reactions that form $^{6}$Li and $^{7}$Li because of the small
cross sections (Meneguzzi et al.~\cite{mar71}).
Thus, the relation between the incident spectrum $j_k(E_{k,0},0)$ and the
spectrum $j_k(E_k,N)$ at depth $N({\rm H}_2)$ in the CSDA is
\be
j_k(E_k,N)=j_k(E_{k,0},0)\frac{\ud E_k}{\ud E_{k,0}}=
j_k(E_{k,0},0)\frac{L_k(E_{k,0})}{L_k(E_k)}\ .
\label{jE0jE}
\ee

%\begin{figure}[!h]
%\begin{center}
%%\includegraphics[height=.40\textheight, width=.85\textwidth]{1794fig8.ps}
%\resizebox{.9\hsize}{!}{\includegraphics{1794fig8.ps}}
%\caption[]{Range $R_{\rm e}(E_{\rm e})$ and $R_{\rm p}(E_{\rm p})$ for electrons and protons
%colliding with H$_2$ ({\em solid} curves), respectively, compared with NIST data ({\em
%circles}) and the results of Cravens \& Dalgarno~(1978, {\em squares});
%the {\em dashed} curve shows the fit by Takayanagi~(1973).}
%\label{Figure8}
%\end{center}
%\end{figure}

\begin{figure}[t]
\begin{center}
\resizebox{.9\hsize}{!}{\includegraphics{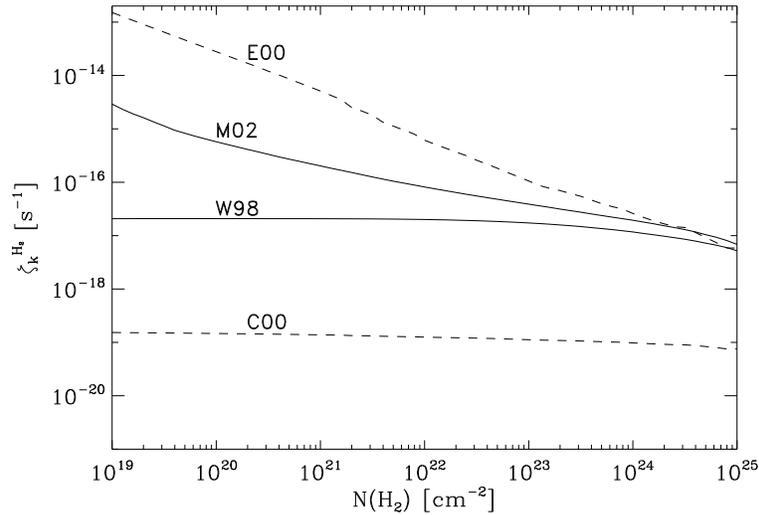}} 
\caption[]{CR ionisation rate $\zeta_k^{{\rm H}_2}$ as a function of the
column density $N({\rm H}_2)$. {\em Solid} curves, contribution of 
CR protons (spectra W98 and M02); {\em dashed} curves, contribution 
of CR electrons (spectra C00 and E00).}
\label{Figure13}
\end{center}
\end{figure}

We are now able to calculate the CR ionisation rate inside a molecular
cloud as a function of the column density, with the attenuated spectra
given by Eq.~(\ref{jE0jE}). 
We compute the CR ionisation rate for
$N({\rm H}_2)$ between $10^{19}$~cm$^{-2}$ and $10^{25}$~cm$^{-2}$, and we
show the results for the four incident LIS spectra in Fig.~\ref{Figure13}.
Since we assume an isotropic CR distribution (see Sect.~\ref{reactions_h2}), 
the ionisation rate calculated for a given $N(\mathrm{H_{2}})$ corresponds to the ionisation
rate at the centre of a spherical cloud with radius $N(\mathrm{H_{2}})$.

As a result of the detailed treatment of CR propagation, the
decrease of the ionisation rate with increasing penetration in the
cloud at column densities in the range $\sim
10^{20}$--$10^{25}$~cm$^{-2}$ is characterised by a power-law behaviour,
rather than exponential attenuation, and can be approximated as
\be
\zeta_k^{{\rm H}_2}\approx 
\zeta_{0,k}\left[\frac{N({\rm H}_2)}{10^{20}~\mbox{cm$^{-2}$}}\right]^{-a}.
\label{zetaNep}
\ee
We have fitted this expression to the numerical results shown in
Fig.~\ref{Figure13}. The coefficients $\zeta_{0,k}$ and $a$ are given in
Table~\ref{tabNep}. The exponential attenuation of the CR ionisation
rate sets in for column densities larger than $\sim 10^{25}$~cm$^{-2}$,
where $\zeta_k^{{\rm H}_2}$ depends essentially on the flux of CR
particles in the high-energy tail of the incident spectrum (above $\sim
0.1$--1~GeV), and directly measurable on the Earth.

\begin{table}[t]
\begin{center}
\caption{Fitting coefficients for Eq.~(\ref{zetaNep}) for CR protons ($p
$, also including heavy nuclei) and electrons ($e$).}
\begin{tabular}{llll}
\hline
$k$ & spectrum   & $\zeta_{0,k}$        & $a$  \\
    &            & [s$^{-1}$]           &      \\
\hline\hline
$p$ & W98        & $2.0\times 10^{-17}$ & $0.021$ \\
$p$ & M02        & $6.8\times 10^{-16}$ & $0.423$ \\
\hline
$e$   & C00      & $1.4\times 10^{-19}$ & $0.040$ \\
$e$   & E00      & $2.6\times 10^{-14}$ & $0.805$ \\
\hline
\end{tabular}
\label{tabNep}
\end{center}
\end{table}

\begin{figure}[t]
\begin{center}
\resizebox{.9\hsize}{!}{\includegraphics{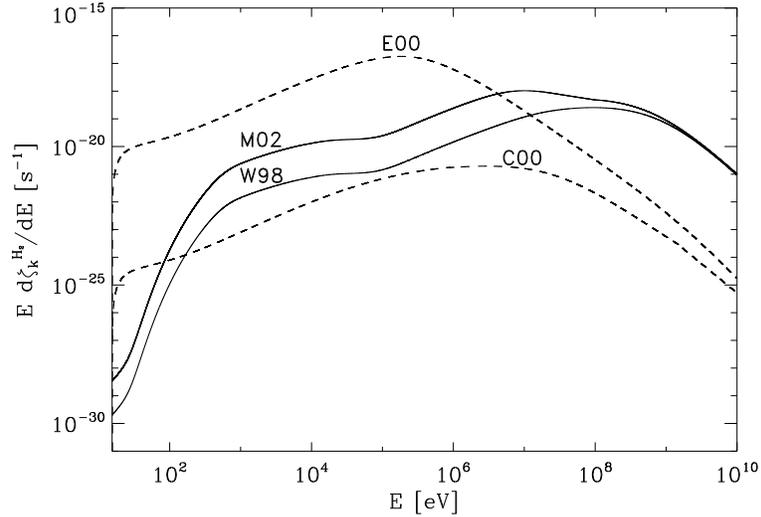}}
\caption[]{Differential contribution to the ionisation rate $E{\rm d}\zeta_k^{{\rm H}_2}/{\rm d}E$ 
per logarithmic interval of kinetic energy, for the four spectra considered
at a depth $N({\rm H}_2)=10^{22}$~cm$^{-2}$ ({\em solid} curves, protons; 
{\em dashed} curves, electrons).}
\label{Figure14}
\end{center}
\end{figure}

It is important to stress that a large contribution to the
ionisation of H$_2$ comes from low-energy protons and electrons
constantly produced (in our steady-state model) by the slowing-down of
more energetic particles loosing energy by interaction with the ambient
H$_2$. In Fig.~\ref{Figure14} we show the differential contribution of CR
protons and electrons to the ionisation rate at a depth of $N({\rm
H}_2)=10^{22}$~cm$^{-2}$, corresponding to the typical column density
of a dense cloud. For protons and heavy nuclei, the bulk of the
ionisation is provided by CR in the range 1~MeV--1~GeV and by a
``shoulder'' in the range 1--100~keV produced by slowed-down protons while, for
electrons, the largest contribution to the ionisation is distributed over
energies in the range 10~keV--10~MeV.
This low-energy tail is produced during the propagation of CR protons and electrons
in the cloud even when the incident spectrum is devoid of low-energy
particles. Thus, the ionisation rate at any
depth in a cloud cannot be calculated by simply removing from the
incident spectrum particles with energies corresponding to ranges below
the assumed depth.

\section{Comparison with observations}
\label{comparison}

\begin{figure}[!h]
\begin{center}
\resizebox{.9\hsize}{!}{\includegraphics{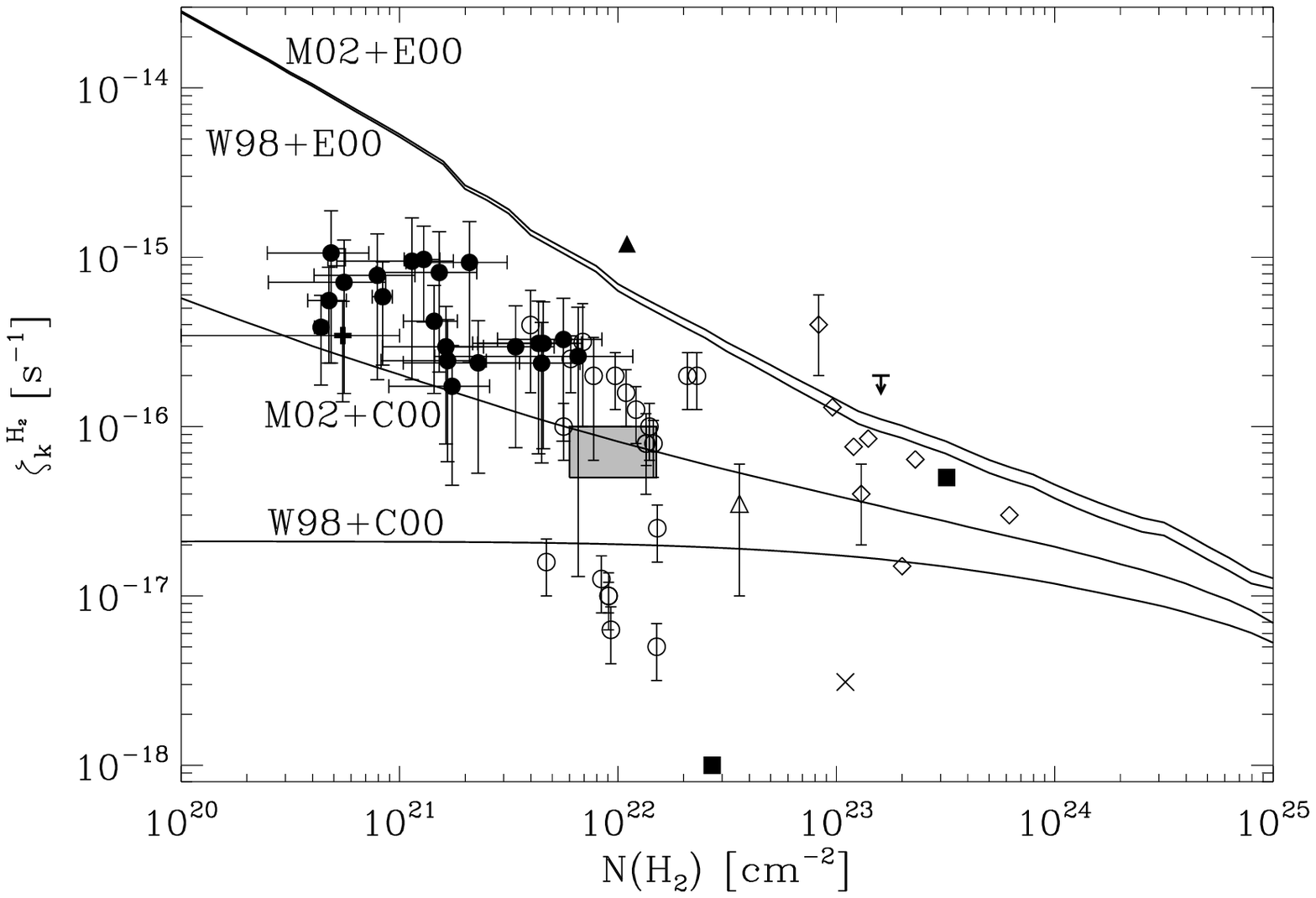}}
\caption[]{Total CR ionisation rate $\zeta^{{\rm H}_2}$ as a function of
$N({\rm H}_2)$ according to our models ({\em solid} curves). Observational
data: 
{\em filled circles}, diffuse clouds (Indriolo et al.~\cite{i12}); 
{\em empty square}, diffuse cloud W49N (Neufeld et al.~\cite{neu10});
{\em empty circles}, dense cores (Caselli et al.~\cite{c98}); 
{\em empty triangle}, prestellar core B68 (Maret \& Bergin~\cite{mb07});
{\em filled squares}, T Tauri disks TW Hya and DM Tau (Ceccarelli et al.~\cite{cec04});
{\em filled triangle}, SNR W51C (Ceccarelli et al.~\cite{cec11});
{\em diamonds}, protostellar envelopes (de Boisanger, Helmich, \& van Dishoeck~\cite{db96},
van der Tak et al.~\cite{vdt00}, van der Tak \& van Dishoeck~\cite{vdtvd00},
Doty et al.~\cite{d02}, and Hezareh et al.~\cite{h08});
{\em cross}, massive star-forming region DR21(OH) (Hezareh et al.~\cite{h08}). 
The {\em filled box} indicates the range of column densities and CR ionisation rates compatible 
with the data analysed by Williams et al.~\cite{w98}.} 
\label{zndati_bw}
\end{center}
\end{figure}

To obtain the total CR ionisation rate in molecular clouds, we sum the
ionisation rates of protons (corrected for heavy nuclei)
and electrons. With two possible spectra
for each component, we obtain four possible profiles of $\zeta^{{\rm
H}_2}$. These are shown in Fig.~\ref{zndati_bw} as a function of $N({\rm
H}_2)$, compared with a compilation of empirical determinations of $\zeta^{{\rm
H}_2}$ in diffuse and dense environments.

The comparison between model results and observational data shown
in Fig.~\ref{zndati_bw} should be taken as indicative and interpreted in a
statistical sense, as also suggested by the large spread of values of
$\zeta^{{\rm H}_2}$ at each value of $N({\rm H}_2)$. First, the {\em
observational} $N({\rm H}_2)$ is the entire column density through the
cloud, whereas the {\em model} $N({\rm H}_2)$ is the column traversed
by CRs incident over the cloud's surface. The exact relation between
the quantities depend on factors like the cloud geometry and
orientation with respect to the line-of-sight, and the variation of CR
ionisation rate with depth within the cloud. In addition, for the
cloud cores of Caselli et al.~\cite{c98} we adopted the H$_2$ column
density estimated by Butner et al.~\cite{bl95} from measurements of
C$^{18}$O multiplied by a factor of 2, to account for depletion of CO
onto grains (Caselli et al.\cite{c98}).  Second, many of
the sight-lines where $\zeta^{{\rm H}_2}$ has been determined in
diffuse clouds may have multiple cloud components, which would reduce
the column density of a single cloud. It is probably safe to conclude that
the {\em observational} column density is an {\em upper limit} to the
column density traversed by CRs incident on each cloud, and therefore
the data shown in Fig.~\ref{zndati_bw} should probably be shifted to the
left by a factor of 2 or so.
At any rate, from the comparison with observational
data, shown in Fig.~\ref{zndati_bw}, we can draw the following
conclusions:

\begin{enumerate}
\item Although the gas column density of the object is by no
means the only parameter controlling the CR ionisation rate, the data
suggest a decreasing trend of $\zeta^{{\rm H}_2}$ with increasing
$N({\rm H}_2)$, compatible with our models M02+C00, W98+E00, W98+C00.
However, the measured values of $\zeta^{{\rm H}_2}$ are very uncertain,
especially in dense environments. Part of the large spread in the
sample of cloud cores may be due to a poor understanding of the
chemistry.
\item The highest values of $\zeta^{{\rm H}_2}$, measured in
diffuse clouds sight lines, could be explained if CR electrons are
characterised by a rising spectrum with decreasing energy. The E00
spectrum represents an extreme example of this kind, and it results in
values of $\zeta^{{\rm H}_2}$ somewhat in excess of the diffuse clouds
observations. The same spectrum accounts simultaneously for the CR
ionisation rates measured in most protostellar envelopes of much higher
column density.  Conversely, a spectrum of protons and heavy nuclei
rising with decreasing energy, like the M02 spectrum, can provide alone
a reasonable lower limit for the CR ionisation rate measured in diffuse
clouds.
\item Without a significant low-energy (below $\sim 100$~MeV)
component of electrons and/or protons and heavy nuclei, it is
impossible to reproduce the large majority of observations. The
combination of the C00 spectrum for electrons with the W98 spectrum for
protons and heavy nuclei clearly fails over the entire range of column
densities.
\end{enumerate}

\section{Effects of magnetic field on CR propagation}
The high values of $\zeta^{{\rm H}_2}$
in the diffuse interstellar gas can be reconciled with the lower values
measured in cloud cores and massive protostellar envelopes by
invoking various mechanisms of CR screening in molecular clouds due to
either self-generated Alfv\'en waves in the plasma~\cite{ss76,hdd78,ps05}
or to focusing and magnetic mirror effects~\cite{cv78,c00,pg11}. 
Besides, a few molecular cloud cores and one dense envelope (Fig.~\ref{zndati_bw}) are
characterised by $\zeta^{{\rm H}_2} \le 10^{-17}$~s$^{-1}$ and probably can only be
explained by invoking the CR suppression mechanisms.
Mirroring and focusing are due to the non-uniformity of the large-scale (mean) component of
the field, whereas diffusion is associated to magnetic field fluctuations on the scale of
the Larmor radius of CR particles.
The relative importance of these processes in the ISM depends on a number of variables
not always well determined (geometry and strength
of the magnetic field, nature and characteristics of turbulence, etc.),
and is unclear whether they can significantly reduce (or enhance) the
CR ionisation of molecular cloud cores. In addition, the damping of
small-scale magnetic fluctuations (e.g. Alfv\'en waves) that affect the
propagation of CRs is strongly dependent on the ionisation fraction of
the medium, which, in turn, is mostly determined by the CRs
themselves.

\begin{figure}[!h]
\begin{center}
\resizebox{.7\hsize}{!}{\includegraphics{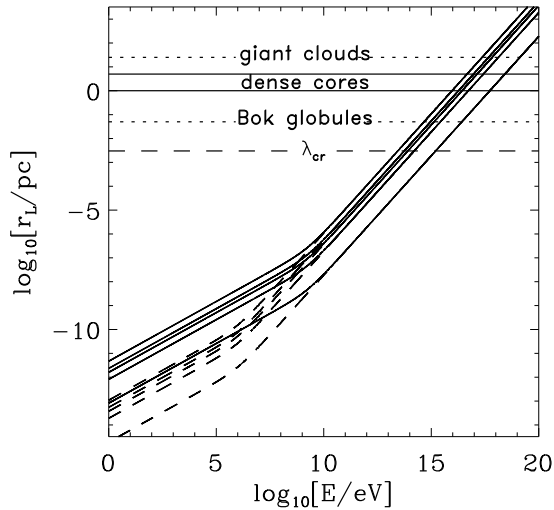}}
\caption[]{Larmor radius as a function of the cosmic-ray energy for 
$\alpha=1¡,\,10¡,\,20¡,\,30¡,\, {\rm and}\ 90¡$.
{\em Solid} and 
{\em dashed lines} represent cosmic-ray protons and electrons, respectively.} 
\label{larmorradius}
\end{center}
\end{figure}

CRs perform an helicoidal motion around the magnetic field lines.
For a uniform magnetic field $B=10\ \mu$G, the Larmor radii, $r_{\rm L}$ of ionising CRs
(CR protons and heavy nuclei with $E \lesssim
1$~GeV/nucleon and CR electrons with $E \lesssim 10$~MeV, see Padovani et al.~\cite{pgg09})
are less than $\sim 10^{-7}$~pc and $\sim 10^{-9}$~pc for protons and electrons,
respectively, many orders of magnitude smaller than the
typical size of Bok globules, dense cores, and giant molecular clouds~\cite{pg11},
see Fig.~\ref{larmorradius}.
In the absence of small-scale
perturbations in the field, we can therefore assume that low-energy CRs
propagate closely following the magnetic field lines.

Perturbations in the forms of magnetohydrodynamic (MHD) waves with
wavelength of the order of the Larmor radius of the particle can
efficiently scatter CRs. The waves can be part of an MHD turbulent
cascade, or can be self-generated by the CRs themselves~\cite{k05}.
However, in a mostly neutral ISM, turbulent MHD
cascades are quenched at scales of roughly the collision mean free path
of ions with neutrals, if the ion-neutral collision rate exceeds the
energy injection rate. Alfv\'en waves are efficiently damped by collision with neutrals
below a critical wavelength $\lambda_{\rm cr}$~\cite{kp69} and, for typical molecular cloud conditions, $\lambda_{\rm cr}\approx3\times10^{-3}$~pc~\cite{pgb08}.
Thus, 
for typical values of the cloud's parameters, only CR particles
with energy larger than a few TeV, and Larmor radii $r_{\rm L} \ge
\lambda_{\rm cr}$ find MHD waves to resonate with. These particles 
however do not contribute significantly to the ionisation of the cloud.

We expect that CRs in the energy range between
100~MeV/nucleon and 1~GeV/nucleon, that provide the bulk of the
ionisation in a cloud core, stream freely through the core without
self-generating MHD waves. We ignore therefore the presence of
self-generated waves in cloud cores in the following, but we
believe that this problem deserves further scrutiny.

\section{Magnetic focusing and mirroring}
\label{focusmirror}
Theoretical models predict that collapsing cloud cores must overcome
the support provided by their magnetic field in order to form stars. In
the process, the competition between gravity pulling inward and
magnetic pressure pushing outward is expected to produce a warped, {\em
hourglass} pattern of the magnetic field. Recently, this scenario has
received support from observations~\cite{gr06,t09} showing a magnetic field geometry
consistent with the formation of solar-type stars, in which
ordered large-scale magnetic fields control the evolution and collapse
of molecular cloud cores~\cite{gg08}. We therefore
adopt the hourglass geometry as the basis of our analysis of CR
penetration into a cloud core.

The effects of magnetic mirroring and focusing in a hourglass geometry
can be simply described following e.g. Desch et al.~\cite{d04}.  A charged
particle travelling in a magnetised medium conserves its kinetic energy
$\gamma mc^2$ and its magnetic moment $\mu=\gamma
mv^2\sin^2\alpha/2B$.  It follows that CRs propagating from the intercloud medium (ICM) to
the cloud's interior must increase $v_\perp^2$ to conserve $\mu$ and
decrease $v_\parallel$ to conserve $|{\bf v}|^2$. Thus, the pitch angle
of the particle must increase from the value $\alpha_{\rm ICM}$ to a
value $\alpha$ as
\be
\frac{\sin^2\alpha}{\sin^2\alpha_{\rm ICM}}=\frac{B}{B_{\rm ICM}}\equiv \chi\,,
\label{pitch}
\ee
where $\chi>1$. Therefore, a CR starting in the ICM with a pitch angle
$\sin\alpha_{\rm ICM} > 1/\chi^{1/2}$ cannot penetrate a region with magnetic
field $B >\chi B_{\rm ICM}$, and will be bounced out ({\em magnetic
mirroring}).  Conversely, the CR flux $j(E)$ in the cloud is increased
by the opening out of the field lines by a factor proportional to the
density of magnetic field lines per unit area ({\em magnetic
focusing}),
\be
j(E)=\chi j_{\raisebox{-3pt}{\tiny\rm{ICM}}}(E)\,.
\ee
The effects of focusing and mirroring depend only on the magnetic
field strength, and are the same for CR protons, electrons, and 
heavy nuclei. 

\section{CR ionisation rate in presence of a magnetic field}
In order to study the effects of the magnetic field on the propagation
of CRs in molecular cores, we adopt the models of Li \& Shu~\cite{ls96}
(see also Galli et al.~\cite{gl99}), for magnetostatic, scale-free,
self-gravitating clouds supported by axially-symmetric hourglass-like
magnetic fields. For simplicity, we consider
models with an isothermal equation of state. These models are
characterised by a value of the non-dimensional mass-to-flux ratio
$\lambda$ defined by
\be
\lambda=2\pi G^{1/2}\frac{M(\Phi)}{\Phi}\,,
\label{def_lambda}
\ee
where $G$ is the gravitational constant, $\Phi$ the magnetic flux, and
$M(\Phi)$ the  mass contained in the flux tube $\Phi$.
For our reference model we choose $\lambda=2.66$, assuming a sound speed $c_{\rm s}=
0.2$~km~s$^{-1}$ and an intercloud magnetic field $B_{\rm ICM}=3~\mu$G. We focus on a flux
tube enclosing a mass $M(\Phi)=1$~M$_{\odot}$, a typical value for a low-mass core.
The following equations are written as a function of the polar angle $\theta$: a CR is 
outside the core for $\theta=0$ and reaches the midplane when $\theta=\pi/2$.

\begin{figure}[!h]
\begin{center}
\resizebox{.7\hsize}{!}{\includegraphics{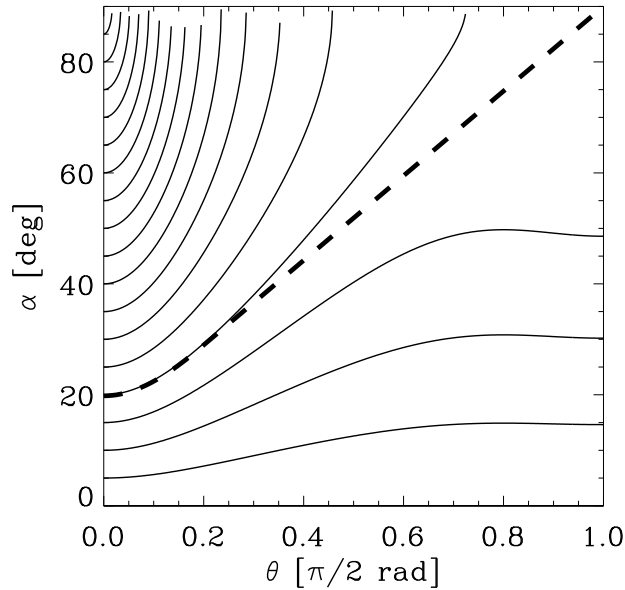}}
\caption[]{Variation of the CR pitch angle as a function of the polar angle for values
of $\alpha_{\rm ICM}$ between 0$^{\circ}$ and 90$^{\circ}$ in steps of 5$^{\circ}$.
The {\em dashed line} represents the critical pitch angle $\alpha_{\rm cr}=20.5^{\circ}$.} 
\label{alfalimit_santcugat}
\end{center}
\end{figure}

Following the assumptions in Sect.~\ref{focusmirror}, 
CRs starting with pitch angles $\alpha_{\rm ICM}$
smaller than a critical value $\alpha_{\rm cr}$ 
are able to reach the cloud's midplane.
When $\alpha_{\rm ICM}>\alpha_{\rm cr}$ then CRs will be pushed back by magnetic mirroring
before reaching the midplane at a position $\theta_{\rm max}(\alpha_{\rm ICM})<\pi/2$.
Inverting this relation, one finds the value of the maximum allowed pitch angle
$\alpha_{\rm ICM,max}(\theta)$ for a CR to reach a given position $\theta$. 
Figure~\ref{alfalimit_santcugat} shows the variation of the pitch angle computed
from Eq.~(\ref{pitch}) assuming the magnetic field profile of the reference 1~M$_{\odot}$
flux tube.

\begin{figure}[!h]
\begin{center}
\includegraphics[scale=0.6]{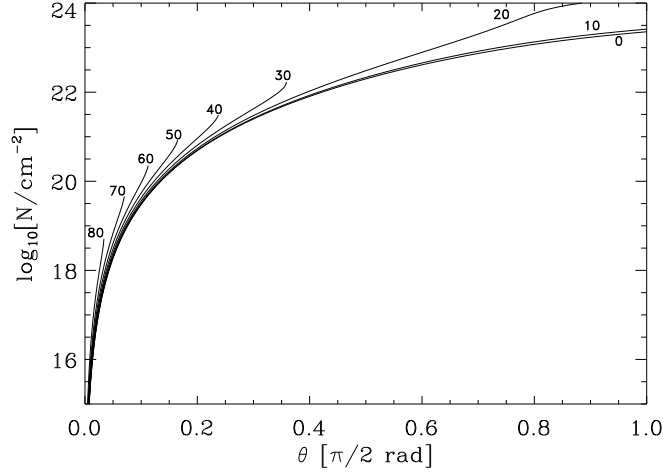}
\caption[]{Column density passed through before reaching the mirror point by
CRs propagating along a field line of the reference flux tube enclosing
1~$M_\odot$, as function of the polar angle $\theta$ for different
values of the initial pitch angle $\alpha_{\rm ICM}$ (labels in
degrees).}
\label{16853fg4}
\end{center}
\end{figure}

The contribution to the CR ionisation rate of H$_{2}$ for CRs entering the core is
\begin{eqnarray}
\zeta^{\rm H_2}(\theta) &=& 2\pi\ \chi(\theta)\int_0^\infty \ud E
\int_0^{\alpha_{\rm ICM, max}(\theta)}
j[E,N(\theta,\alpha_{\rm ICM})]\nonumber\\
&&\times [1+\phi(E)]\ \sigma^{\rm ion}(E)\ 
\sin\alpha_{\rm ICM}\ \ud\alpha_{\rm ICM}\,,
\label{ion1}
\end{eqnarray}
where $N(\theta,\alpha_{\rm ICM})$ is the column density of H$_2$ 
into the core (with
$N=0$ at $\theta=0$), $\phi(E)$ is given by Eq.~(\ref{secondel}), 
$\sigma^{\rm ion}(E)$ is the ionisation cross section of H$_2$  
(see Sect.~\ref{reactions_h2}), and the column density passed through by a CR propagating
along a magnetic field line is
\be
\label{colden}
N(\theta,\alpha_{\rm ICM})=\frac{1}{\mu\,m_{\rm H}}\int\rho\,
\frac{\ud\ell}{\cos\alpha},
\ee
where $\mu=2.8$ is the molecular weight, $\ud\ell=[\ud
r^2+(r\ud\theta)^2]^{1/2}$ is the element of magnetic 
field line, and the factor
$1/\cos\alpha$ accounts for the increase of the actual path length of a
CR performing a helicoidal trajectory around a magnetic field with
respect to the displacement along the field line. 
Fig.~\ref{16853fg4} shows the column density passed through by the CR before
reaching the mirror point.
As a first approximation one can assume that CRs
coming from the ICM and travelling towards the cloud's midplane experience a similar
increase in column density, independently on the initial pitch angle $\alpha_{\rm ICM}$,
the latter mainly determining the value of the column density at which the CRs are
pushed out by magnetic mirroring. For this reason, we assume for all CRs the column
density profile corresponding to $\alpha_{\rm ICM}=0$, but truncated at increasingly
larger values depending on the initial pitch angle $\alpha_{\rm ICM}$,
\be
N(\theta,\alpha_{\rm ICM}) \approx N(\theta,0) 
~~~\mbox{if~~~ $0<\theta<\theta_{\rm max}(\alpha_{\rm ICM})$}\,.
\ee

Adopting the CSDA (see Sect.~\ref{propagation}), the total CR ionisation rate,
see details in Padovani \& Galli~\cite{pg11}, is then given by two contributions
due to CRs entering the core from the upper side (subscript $+$) and the lower side
(subscript $-$) of the core
\be\label{zetatot}
\zeta^{\rm H_2}(\theta)=\zeta_+^{\rm H_2}(\theta)+\zeta_-^{\rm H_2}(\theta)\,,
\ee
with
\begin{eqnarray}
\zeta_+^{\rm H_2}(\theta)&=&\varphi_+(\theta)\ 
\zeta_0^{\rm H_2}[N_0(\theta)]\nonumber\\
\zeta_-^{\rm H_2}(\theta)&=& \varphi_-(\theta)\ 
\zeta_0^{\rm H_2}[2N_0(\pi/2)-N_0(\theta)]\,.
\end{eqnarray}
$\varphi_{+}(\theta)$ and $\varphi_{-}(\theta)$ are factors accounting for magnetic effects
\begin{eqnarray}
\varphi_+(\theta)&=&\frac{\chi(\theta)}{2}[1-\cos\alpha_{\rm ICM,max}(\theta)]\nonumber\\
\varphi_-(\theta)&=&\frac{\chi(\theta)}{2}(1-\cos\alpha_{\rm cr})\,
\end{eqnarray}
and $\zeta_{0}^{\rm H_{2}}[N_{0}(\theta)]$ is the CR ionisation rate in the non-magnetic
case, see Eqs.~(\ref{zetatot_h2}), (\ref{zetatot_he}), and (\ref{zetatot_e}).

\begin{figure}[!h]
\begin{center}
\resizebox{\hsize}{!}{\includegraphics{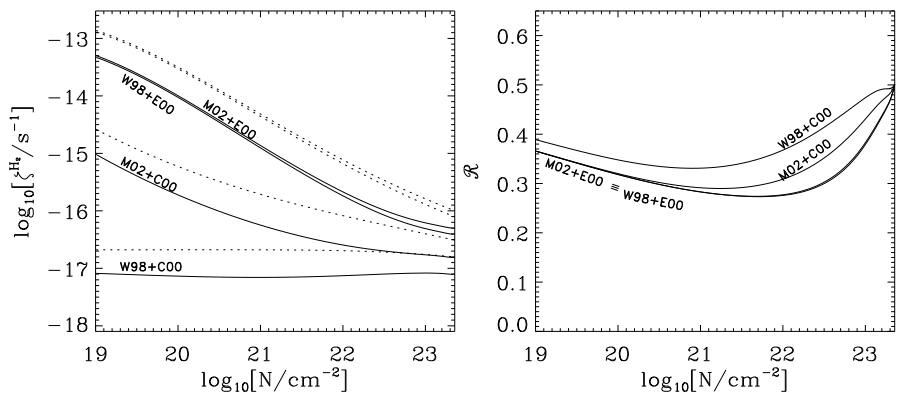}}
\caption{{\em Left panel}: comparison between the CR ionisation rate with and without the
effects of magnetic field ({\em solid} and {\em dotted lines},
respectively) as function of the column density.
{\em Right panel}: ratio between the ionisation rates in the magnetic and non-magnetic case. The curves are labelled as in Fig.~\ref{zndati_bw}.}
\label{propagazione_ottavo5_santcugat}
\end{center}
\end{figure}

Mirroring and focusing have opposite effects of comparable magnitude on the ionisation
rate, both becoming more and more important approaching the core's midplane where the 
field is stronger. However, as shown in the left panel of
Fig.~\ref{propagazione_ottavo5_santcugat}, magnetic mirroring always
reduces the CR ionisation rate more than magnetic focusing can increase it, the total
effect being a net reduction of $\zeta^{\rm H_{2}}$ by a factor between 2 and 3.
Notice that the maximum effect of the magnetic field on $\zeta^{\rm H_{2}}$ is obtained at
an intermediate position corresponding to column densities of $10^{21}-10^{22}$~cm$^{-2}$
where the reduction factor ${\cal R}=\zeta^{{\rm H}_{2}}/\zeta^{{\rm H}_{2}}_{0}$ is
about 0.3 (see right panel of Fig.~\ref{propagazione_ottavo5_santcugat},
while the value at the core's midplane is independent on the assumed CR spectrum,
a result also obtained by Desch et al.~\cite{d04}.

We explored the parameter space, investigating the variation of the relevant
quantities of the problem as a function of the enclosed mass for our reference core.
Flux tubes enclosing smaller masses intersect the
midplane at smaller equatorial radii ($r_{\rm eq}$) and are characterised by larger values
of the magnetic field and density. As a consequence, focusing becomes
more important, since $\chi$ increases, but also mirroring becomes more
severe, since $\alpha_{\rm cr}$ decreases. The net effect is a stronger
reduction of $\zeta^{\rm H_2}$ in the innermost regions of the core as
compared to the envelope (see Table~\ref{tab:BvsM}). As the field
strength increases approaching the central singularity of the model,
$\chi \rightarrow \infty$ and the reduction factor 
of the CR ionisation rate in the core's midplane
approaches the asymptotic value ${\cal R}(\pi/2)=1/2$.
Conversely, for flux tubes enclosing larger masses, the field strength
approaches the ICM value, and the density decreases to zero.
Therefore both focusing and mirroring become weaker, and $\alpha_{\rm
cr}$ approaches $90^\circ$, as shown by Table~\ref{tab:BvsM}.
As expected, for increasing values of $M(\Phi)$, $\zeta^{{\rm H_2}}$
approaches the value of the non-magnetic case, because the magnetic
field strength decreases away from the centre of the core, 
approaching the ICM value.

\begin{table}[!h]
\caption{Values of the parameters described in the text as function of the 
mass, $M(\Phi)$, contained within flux tube $\Phi$.}
\begin{center}
\begin{tabular}{cccccc}
\hline\hline
$M(\Phi)$ & $r_{\rm eq}$ & $N_0(\pi/2)$ & $\alpha_{\rm cr}$ 
& $\chi(\pi/2)$ & ${\cal R}(\pi/2)$\\
$(M_\odot)$ & (pc) & (10$^{23}$ cm$^{-2}$) &  &  & \\
\hline
0.5 & 0.018 & 4.54 & 14.8$^\circ$ & 15.347 &  0.508\\
1   & 0.036 & 2.27 & 20.5$^\circ$ & 8.174  &  0.516\\
5   & 0.180 & 0.45 & 39.9$^\circ$ & 2.435  &  0.566\\
10  & 0.360 & 0.23 & 49.7$^\circ$ & 1.717  &  0.608\\
50  & 1.802 & 0.05 & 69.3$^\circ$ & 1.143  &  0.739\\
100 & 3.604 & 0.02 & 75.0$^\circ$ & 1.072  &  0.794\\
\hline
\end{tabular}
\end{center}
\label{tab:BvsM}
\end{table}%

\begin{figure}[!h]
\begin{center}
\includegraphics[scale=0.55]{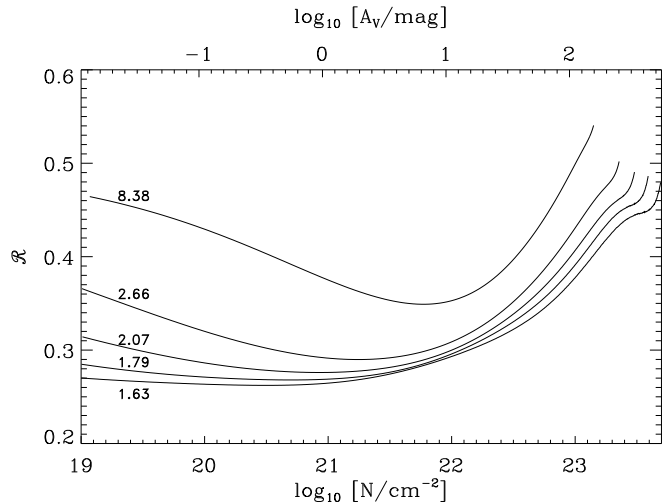}
\caption{Reduction factor ${\cal R}$ between the CR ionisation rates in the magnetic and
non-magnetic case for the case of M02+C00 spectrum (see
Figure~\ref{zndati_bw} for the label). The curves are computed for 
a flux tube containing 1~$M_\odot$ and for different values of
$\lambda=8.38$, 2.66, 2.07, 1.79, 1.63. The upper scale shows the extinction
through the cloud obtained from $A_V/{\rm mag}=N/10^{21}~{\rm cm}^{-2}$.}
\label{16853fg7}
\end{center}
\end{figure}

For cores with strong magnetic support (lower values of $\lambda$), the
equatorial squeezing of the field lines is stronger, and the lines
reach more internal regions of the core where the density is higher.
As a consequence, $\alpha_{\rm cr}$ decreases, the mirroring effect
becomes stronger, and a smaller fraction of CRs can penetrate the
cloud.  In Figure~\ref{16853fg7} we show the reduction factor ${\cal R}$ as a
function of column density for decreasing values of $\lambda$.  For
simplicity, we have considered only the combination of the proton
spectrum M02 and the electron spectrum C00 (the results obtained with
the other spectra are similar).  As the Figure shows, the
reduction of $\zeta^{\rm H_2}$ is larger in cores with larger magnetic
support, due to the increase in the field strength and concentration of
field lines. The reduction is a factor $\sim 4$ for the outer regions
of cores with $\lambda=1.63$, the lowest value of mass-to-flux 
ratio considered in our models.

For $\lambda\rightarrow 1$, the density distribution becomes more and
more flattened, the core assumes the shape of a thin disk, and the
column density from the ICM to the core's midplane becomes larger. 
For these magnetically dominated 
disk-like configurations, the reduction of the CR ionisation rate 
approaches the asymptotic value 1/2.

\section{Conclusions}
\label{conclusions}
The comparison between our models and the observational data available
for diffuse clouds, dense cores and massive protostellar envelopes
indicates that good agreement between theory and observations can be
obtained for the CR ionisation rate of the ISM by including CR
electrons with an energy spectrum increasing towards low energies, as
also suggested by Webber~\cite{w98}.
Despite the observational uncertainties due to the uncertainty in the CR
spectrum at energies below $\sim 1$~GeV and the uncertainties in the
empirically determined values of $\zeta^{{\rm H}_2}$ in diffuse and
dense molecular clouds, several important conclusions clearly emerge from
our study:
\begin{enumerate}
\item Values of $\zeta^{{\rm H}_2}$ measured in diffuse clouds
are greater on average by an order of magnitude than those ones
measured in dense molecular clouds. If confirmed, these data imply the
presence of a CR proton and/or CR electron spectrum which increases at
low energies.

\item Values of $\zeta^{{\rm H}_2}$ measured in dense molecular
clouds span a range of about two orders of magnitude and are subject to
considerable uncertainty. It is difficult to establish how much of the
observed spread is due to variations in the CR ionisation rate. It is
likely that in dense clouds the effects of magnetic fields on the
propagation of CR particles cannot be neglected. In addition, it might
be necessary to take into account the density distribution inside each
cloud.

\item The values of $\zeta^{{\rm H}_2}$ measured in massive
protostellar envelopes are somewhat higher than the predictions of our
models at the corresponding column densities. This seems to suggest the
presence of further ionisation sources in these objects, as, for
example, X-ray emission from the young stellar objects.

\item $\zeta^{{\rm H}_2}$ in a magnetised core is always reduced
with respect to its non-magnetic value, by a factor depending on the core's mass-to-flux
ratio ($\lambda$) 
and the amount of mass contained in the flux tube considered. The reduction is
less severe for flux tubes enclosing larger masses and for larger values of $\lambda$.
Thus, the values of $\zeta^{\rm H_2}$ derived for dense cores and 
globules~\cite{c98,wal98,mb07} probably underestimate the ``external'' (i.e. intercloud) CR
ionisation rate by a factor of $\sim 3-4$, thus alleviating the
discrepancy with measurements of $\zeta^{\rm H_{2}}$ in diffuse clouds.

\end{enumerate}


\begin{thebibliography}{99.}%

\bibitem{bd77}
Black, J.~H. \& Dalgarno, A.\ 1977, ApJS, 34, 405

\bibitem{bhd78}
Black, J.~H., Hartquist, T.~W. \& Dalgarno, A.\ 1978, ApJ, 224, 448

\bibitem{bl95} 
Butner, H.~M., Lada, E.~A., Loren, R.~B.\ 1995, ApJ, 448, 207

\bibitem{cb04}
Casadei, D. \& Bindi, V.\ 2004, ApJ, 612, 262 

\bibitem{c98} 
Caselli, P., Walmsley, C.~M., Terzieva, R., et al.\ 1998, ApJ, 499, 234

\bibitem{cec04}
Ceccarelli, C., Dominik, C., Lefloch, B., et al.\ 2004, ApJ, 607, L51

\bibitem{cec11}
Ceccarelli, C., Hily-Blant, P., Montmerle, T., et al.\ 2011, ApJL, 740, L4

\bibitem{cpa92}
Cecchi-Pestellini, C. \& Aiello, S.\ 1992, MNRAS, 258, 125

\bibitem{cv78}
Cesarski, C.~J. \& V\"olk, H.~J.\ 1978, A\&A, 70, 367

\bibitem{c00}
Chandran, B.~D.~G.\ 2000, ApJ, 529, 513

\bibitem{cvd75}
Cravens, T.~E., Victor \& G.~A., Dalgarno, A.\ 1975, Plan. Space Sci., 23, 1059 

\bibitem{cd78} 
Cravens, T.~E. \& Dalgarno, A.\ 1978, ApJ, 219, 750

\bibitem{dyl99} 
Dalgarno, A., Yan, M. \& Liu, W.\ 1999, ApJS, 125, 237 

\bibitem{db96}
de Boisanger, C., Helmich, F.~P. \& van Dishoeck, E. F.\ 1996, ApJ, 463, 181

\bibitem{d04} 
Desch, S.~J., Connolly, H.~C., Jr. \& Srinivasan, G.\ 2004, ApJ, 602, 528 

\bibitem{d02} 
Doty, S.~D., van Dishoeck, E.~F., van der Tak, F.~F.~S., et al.\ 2002, A\&A, 389, 446

\bibitem{fwl96}
Federman, S.~R., Weber, J. \& Lambert, D.~L.\ 1996, ApJ, 463, 181

\bibitem{gab07} 
Gabici, S., Aharonian, F.~A. \& Blasi, P.\ 2007, Ap\&SS, 309, 365 

\bibitem{gl99}
Galli, D., Lizano, S., Li, Z.-Y., et al.\ 1999, ApJ, 521, 630

\bibitem{g99}
Geballe, T.~R., McCall, B.~J., Hinkle, K.~H., et al.\ 1999, ApJ, 510, 251 

\bibitem{gl73b}
Glassgold, A.~E. \& Langer, W.~D.\ 1973, ApJ, 186, 859

\bibitem{gr06}
Girart, J.~M., Rao, R. \& Marrone, D.~P.
2006, Science, 5788, 812

\bibitem{gl74}
Glassgold, A.~E. \& Langer, W.~D.\ 1974, ApJ, 193, 73

\bibitem{gg08}
Gon\c calves, J., Galli, D. \& Girart, J.~M.\ 2008, A\&A, 490, L39

\bibitem{hbd78}
Hartquist, T.~W., Black, J.~H. \& Dalgarno, A.\ 1978, MNRAS, 185, 643

\bibitem{hdd78}
Hartquist, T.~W., Doyle, H.~T. \& Dalgarno, A.\ 1978, A\&A, 68, 65

\bibitem{hnt61}
Hayakawa, S., Nishimura, S. \& Takayanagi, T.\ 1961, PASJ, 13, 184 

\bibitem{ht05}
Heiles, C. \& Troland, T.~H.\ 2005, ApJ, 624, 773 C

\bibitem{h08} 
Hezareh, T., Houde, M., McCoey, C., et al.\ 2008, ApJ, 684, 1221 

\bibitem{k05}
Kulsrud, R.~M.\ 2005, Plasma physics for astrophysics, Princeton University Press

\bibitem{kp69} 
Kulsrud, R. \& Pearce, W.~P.\ 1969, ApJ, 156, 445 

\bibitem{i12}
Indriolo, N. \& McCall, B.~J.\ 2012, ApJ, 745, 91

\bibitem{ls96}
Li, Z.-Y. \& Shu, F.~H.\ 1996, ApJ, 472, 211

\bibitem{mb07} 
Maret, S. \& Bergin, E.~A.\ 2007, ApJ, 664, 956 

\bibitem{mar71}
Meneguzzi, M., Adouze, J. \& Reeves, H. 1971, A\&A, 15, 377

\bibitem{m04} 
Mewaldt, R.~A., Wiedenbeck, M.~E., Scott, L.~M., et al.\ 2004,
Physics of the Outer Heliosphere, AIP Conference Proceedings, vol.~719, p.~127

\bibitem{mcc98} 
McCall, B.~J., Geballe, T.~R., Hinke, K.~H., et al.\ 1998, Science, 279, 1910

\bibitem{mcc03} 
McCall, B.~J., Huneycutt, A.~J., Saykally, R.~J., et al.\ 2003, Nature, 422, 500

\bibitem{m02} 
Moskalenko, I.~V., Strong, A.~W., Ormes, J.~F. et al.\ 2002, ApJ, 565, 280

\bibitem{neu10}
Neufeld, D.~A., Sonnentrucker, P., Phillips, T.~G., et al.\ 2010, A\&A, 518, L108

\bibitem{ps05}
Padoan, P. \& Scalo, 2005, ApJ, 624, L97

\bibitem{pgg09}
Padovani, M., Galli, D. \& Glassgold, A.~E.\ 2009, A\&A, 501, 619

\bibitem{pg11}
Padovani, M. \& Galli, D.\ 2011, A\&A, 503, A109

\bibitem{pav90} 
Phelps, A.~V.\ 1990, J. Phys. Chem. Ref. Data, 19, 3 

\bibitem{pgb08} 
Pinto, C., Galli, D. \& Bacciotti, F.\ 2008, A\&A, 484, 1 

\bibitem{p95} 
Potgieter, M.~S.\ 1995, Adv. Space Res., 16 (9), 191

\bibitem{rl75} 
Ramaty, R., \& Lingenfelter, R.~E.\ 1975, Solar Gamma-, X-, and EUV Radiation, 68, 363 

\bibitem{rl96} 
Ramaty, R., Kozlovsky, B., \& Lingenfelter, R.~E.\ 1996, ApJ, 456, 525 

\bibitem{s02} 
Schlickeiser, R.\ 2002, in Cosmic Ray Astrophysics, (Springer: Berlin)

\bibitem{ss76}
Skilling, J. \& Strong, A.~W.\ 1976, A\&A, 53, 253 

\bibitem{sng71}
Swartz, W.~E., Nisbet, J.~S. \& Green, A.~E.~S.\ 1971, J. Geophys. Res., 76, 8425

\bibitem{st68} 
Spitzer, L. \& Tomasko, M.~G.\ 1968, ApJ, 152, 971

\bibitem{smr00} 
Strong, A.~W., Moskalenko, I.~V. \& Reimer, O.\ 2000, ApJ, 537, 763 

\bibitem{t73} 
Takayanagi, M.\ 1973, PASJ, 25, 327

\bibitem{t09} 
Tang, Y.-W., Ho, P.~T.~P., Koch, P.~M., et al.\ 2009, ApJ, 700, 251 

\bibitem{un81} 
Umebayashi, T. \& Nakano, T.\ 1981, PASJ, 33, 617

\bibitem{vdtvd00}
van der Tak, F.~F.~S. \& van Dishoeck, E.~F.\ 2000, A\&A, 358, L79

\bibitem{vdt00} 
van der Tak, F.~F.~S. \& van Dishoeck, E.~F. \& Evans, N.~J., II et al.\ 2000, ApJ, 537, 283 

\bibitem{vdb86}
van Dishoeck, E.~F. \& Black, J.~H.\ 1986, ApJS, 62, 109

\bibitem{w87}
Webber, W.~R.\ 1987, A\&A, 179, 277 

\bibitem{w98} 
Webber, W.~R.\ 1998, ApJ, 506, 329 

\bibitem{wal98}
Williams, J.~P., Bergin, E.~A., Caselli, P. et al.\ 1998, ApJ, 503, 689 

\end{thebibliography}
\end{document}